\documentclass[a4paper,11pt]{article}
\pdfoutput=1 

\usepackage{jcappub} 
\usepackage{aas_macros} 

\usepackage[T1]{fontenc} 
\usepackage{mathtools}

\usepackage{array}
\newcolumntype{P}[1]{>{\centering\arraybackslash}p{#1}}
\newcolumntype{M}[1]{>{\centering\arraybackslash}m{#1}}
\usepackage{graphicx}
\usepackage[dvipsnames]{xcolor}
\usepackage{longtable}
\usepackage{url}
\usepackage{verbatim}
\usepackage[utf8]{inputenc}
\usepackage{makecell}
\usepackage{comment}
\usepackage{cleveref}
\crefname{section}{§}{§§}
\usepackage[T1]{fontenc} 
\graphicspath{ {Figures/} }

\title{\boldmath Delayed teraelectronvolt emission from GRB 980425/SN 1998bw and the origin of ultra-high-energy cosmic rays}

\author[a,b,c]{Nestor Mirabal}

\affiliation[a]{Mail Code 661, Astroparticle Physics Laboratory,
NASA Goddard Space Flight Center, Greenbelt, MD 20771,
USA}
\affiliation[b]{University of Maryland, Baltimore County, MD 21250, USA}
\affiliation[c]{Center for Research and Exploration in Space Science and Technology, NASA Goddard Space Flight Center, Greenbelt, MD 20771}

\emailAdd{nestor.r.mirabalbarrios@nasa.gov}

\abstract{
The origin of ultra-high-energy cosmic rays (UHECRs, E $> 10^{18}$ eV) is one of the great mysteries of modern astrophysics. It has been suggested that UHECRs could be accelerated in 
gamma-ray bursts (GRBs) and engine-driven supernovae (SNe). Here we report the discovery of a 1.4 teraelectronvolt (TeV) photon  offset 0.97$^{\circ}$ from the site of the nearby (36.9 megaparsecs) GRB 980425/SN 1998bw explosion.
The Large Area Telescope (LAT) on board the Fermi 
Gamma-ray Space Telescope detected the TeV emission on 17 November 2018, more than 20 years after the original GRB 980425/SN 1998bw trigger. TeV detections at high Galactic latitudes by the LAT are extremely rare, with an average of 6 events per year. We propose that the delayed TeV emission is consistent with ultra-high-energy cosmic rays and/or electron-positron pairs from GRB 980425/SN 1998bw being deflected by the intergalactic magnetic field (IGMF) and subsequently cascading into secondary gamma rays. Based on the arrival time delay of the TeV emission, we estimate an IGMF strength  of order $B \simeq 10^{-12}$--$10^{-13}$ Gauss. This result supports the possibility of UHECR acceleration in GRB 980425/SN 1998bw  and suggests that most detected UHECRs are produced in  local GRB/SNe within 200 Mpc. In addition, secondary photons from UHECRs out to 0.9--31  Gpc may also offer an explanation for extragalactic background photons  with energies 
$\geq 1$ TeV detected by the Fermi LAT.}
\begin{document}
\maketitle
\flushbottom

\section{Introduction}
 GRBs have long been proposed as the source of UHECRs 
\cite{waxman,vietri}. A scenario in which UHECRs are produced by GRBs should also generate cascades of secondary particles, gigaelectronvolt (GeV) and TeV photons as UHECRs propagate from their sources to the Earth \cite{waxman1,ferrigno2005high,armengaud2006secondary,gabici2007gamma}. Because we live in a magnetized Universe, magnetic fields deflect the trajectory of charged particles 
causing propagated delays on the arrival times of secondary photons.  
It is estimated that a significant fraction of gamma-ray photons  produced in such cascades would reach our detectors with time delays on the order of  years after the GRB detection, depending on the strength of the IGMF involved and the pervasiveness of  magnetic voids \cite{waxman1}.  

The first direct evidence that  long  ($\geq$ 2 s) GRBs accompany some core-collapse type Ibc SNe was found on 25 April 1998  with the discovery of the nearby GRB980425/SN 1998bw at a redshift of $z=0.0085$ \cite{soffita,tinney,galama,kulkarni}. 
Located at a distance of 36.9 megaparsecs (Mpc), GRB 980425/SN 1998bw lies well within the Greisen–Zatsepin–Kuzmin (GZK) distance horizon \cite{greisen,zatsepin}
and it remains the nearest known GRB/SN. Despite its proximity, GRB 980425 was not tremendously energetic, but it was accompanied by a Type Ibc supernova, SN 1998bw, whose unique characteristics provided the observational link between GRBs and stellar explosions. If  UHECRs were accelerated in the  GRB/progenitor of GRB 980425 \cite{waxman,2022MNRAS.514.6011B} or as part of the emerging supernova SN 1998bw  \cite{2006ApJ...651L...5M,wang,2011NatCo...2..175C}, 
secondary photons produced in electromagnetic cascades may follow much later. 
Here, we report a search for this delayed TeV emission near the GRB 980425/SN 1998bw position with the Fermi LAT.

\section{Results}
The Fermi LAT is a pair-conversion detector covering the energy range from about 20 megaelectronvolt (MeV) to more than 2 TeV \cite{atwood}. It has been operated nearly uninterrupted in an all-sky survey mode since August 2008. 
Using more than 14 years of Fermi-LAT data collected between 4 August 2008
and 20 September 2022, we searched for delayed $\geq 1$ TeV emission around the position of GRB 980425/SN 1998bw. A detailed study of isotropic gamma-ray emission at energies $>$ 100 GeV  has been presented elsewhere \cite{raniere}. Our LAT analysis uses standard methods and we only apply two cuts to the LAT event list with the Fermitools.  Using $gtselect$
we selected front/back Pass 8 events belonging to the P8R3\_SOURCEVETO
class (evclass = 2048, evtype = 3), which is similar to  
the P8R3\_SOURCE selection but includes an additional cut on the 
Anti-Coincidence Detector (ACD) tile signal that depends on whether or not there is one non-hit tracker silicon strip detector (SSD)  between the head of the track and the ACD \cite{bruel}.  A zenith angle cut of 90$^{\circ}$ together with a $gtmktime$ cut of DATA\_QUAL$>$0 \&\& LAT\_CONFIG==1  was also applied. For $\geq$ 1 TeV incident photons, the 68\% Fermi-LAT containment angle is less than 0.1$^{\circ}$, but the effective area drops off in the TeV energy range \cite{atwood}. 
 
\begin{figure}[t]%
\centering
\includegraphics[width=1.0\textwidth]{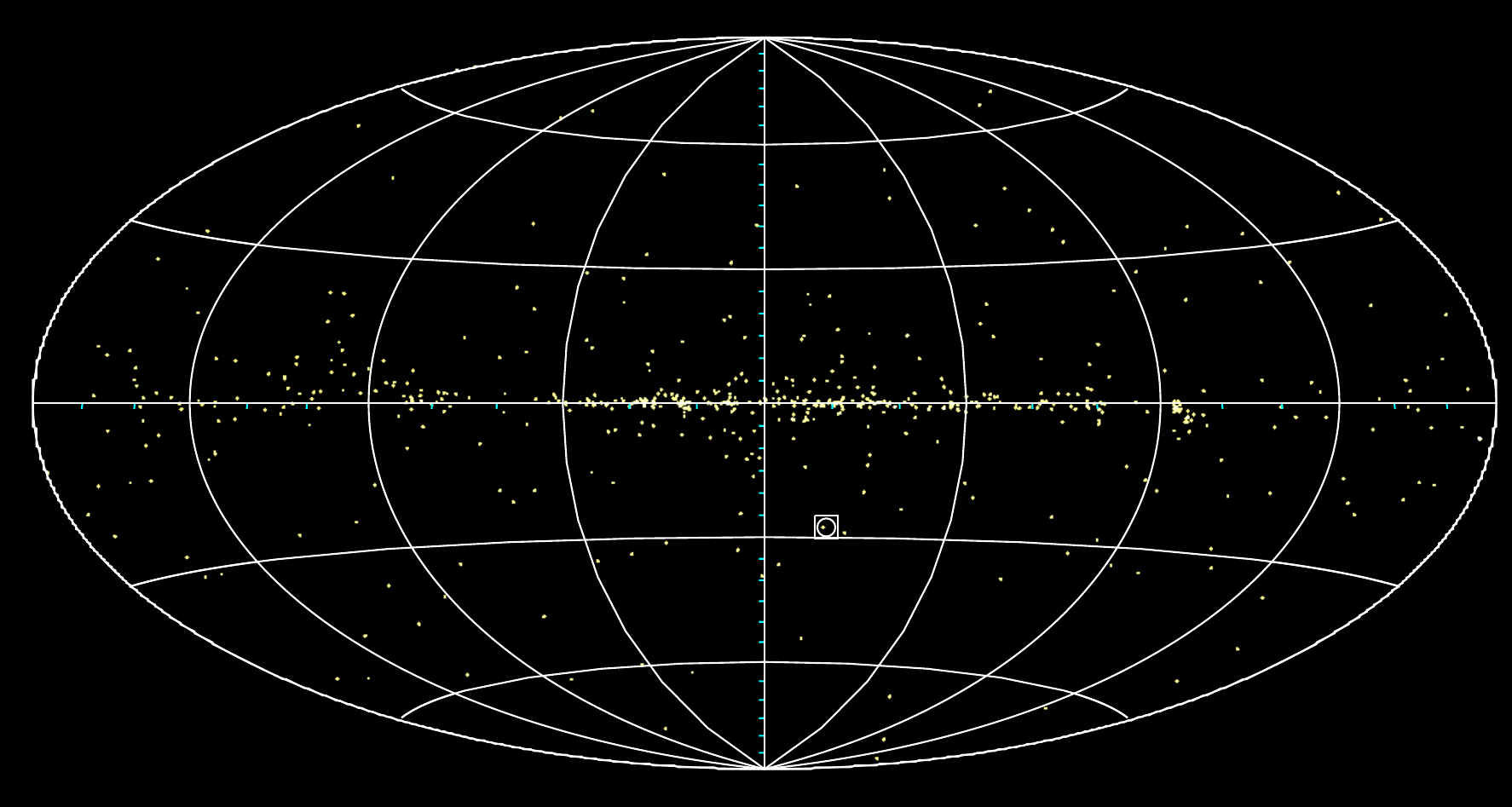}
\caption{Fermi LAT all-sky map
for energies E $\geq$ 1 TeV based on events collected during the period 4 August 2008--20 September 2022
The white square/circle encloses the position of GRB 980425/SN 1998bw and the 1.4-TeV photon.
}\label{fig1}
\end{figure}

 Offset 0.97$^{\circ}$ from the  GRB 980425/SN 1998bw position derived by \cite{fynbo}, 
we find a 1.4 TeV photon at right ascension (RA) = 294.2667$^{\circ}$ and declination (dec.) = -51.92068$^{\circ}$ (J2000); Galactic coordinates ($\ell,b=346^{\circ}.0814,-27^{\circ}$.87171) detected on 17 November  2018  09:32:35.630 UTC.  Figure \ref{fig1} shows the position of the 1.4-TeV photon and GRB 980425/SN 1998bw in the gamma-ray sky. The next two nearest photons from the  GRB 980425/SN 1998bw position lie at 4.0 $^{\circ}$ and 10.6 $^{\circ}$ respectively. 
Cross matching the $\geq$ 1 TeV LAT photon positions  with the 4FGL-DR3 catalog \cite{4fgl,4fgldr3} produced 10 matches with known, catalogued gamma-ray sources assuming a matching radius $r_{m}$ = 0.12$^{\circ}$, see Table~\ref{tab1}. The matching radius  was fixed at the maximum angular separation measured from the  single largest  source contributor at $\geq 1$ TeV energies (6 matched photons), which  is Markarian 421 (Table~\ref{tab1}).

Up to 17 November 2018, there are only 59 LAT photons in the energy range 1.0--3.8 TeV at Galactic latitudes $b \geq \vert 25^{\circ} \vert$.
Assuming 10 photons matched to individual 4FGL sources, this leaves us with 49 unmatched $\geq$ 1 TeV photons. 
The Poisson probability $p$, or the probability that one or more of these 49 unmatched photons would fall within a 1-degree circle of GRB 980425/SN 1998bw as a function of the expectation value $\lambda$, is given by 

\begin{equation}
p=1-e^{-\lambda} = 1 - e^{-r_1\times r_2}
\end{equation}

\noindent 
where $r_1 = 0.0028$ TeV photons deg$^{-2}$ stands for 49 unmatched photons  over 1.7$\pi$ steradians at Galactic latitudes $b \geq \vert 25^{\circ} \vert$ covered by the Fermi LAT, and a 1-degree circular region yields $r_2 = 3.1416$~deg$^{2}$. This results in a Poisson probability of having one or more chance coincidences $p = 0.0088$.
In other words, the probability of a chance association between GRB 980425/SN 1998bw and the 1.4-TeV photon is estimated to be about 0.88\%.

\begin{longtable}{cccc}
\label{tab1}
    \small
RA $(^{\circ})$ & dec.$(^{\circ})$ & Separation $(^{\circ})$  & 4FGL Match\\
\hline
57.33071    & -11.96453   & 0.037  & 1ES 0347-121  \\
80.75072    & -36.45587   & 0.017 & PKS 0521-36  \\
157.8484    & -26.09274   & 0.117  & NVSS J103137-260 \\
166.1172    & 38.21111   & 0.004  & Markarian 421  \\
166.1227    & 38.1954   & 0.012  & Markarian 421  \\
166.1717   & 38.15629   & 0.066  & Markarian 421  \\
166.1795    & 38.25542   & 0.068  & Markarian 421  \\
166.202    & 38.23114   & 0.070  & Markarian 421  \\
166.2408   & 38.13483   & 0.120  & Markarian 421  \\
253.3445   & 39.71316   & 0.110  & Markarian 501 \\
  \hline
    \caption{The entire list of TeV LAT events above 1 TeV (P8R3\_SOURCEVETO) at Galactic latitudes $b \geq \vert 25^{\circ} \vert$ up to 17 November 2018 is available at the following \href{https://zenodo.org/record/7230266\#.Y1Fpd-zMKEs}{link}}
\end{longtable}

One must bear in mind that magnetic deflections are actually expected at various points during the cascade development, thus the probability of a positional chance association should be formally lower. 

For completeness, we searched for potential multi-wavelength counterparts near the TeV emission in the astronomical archives of the High Energy Astrophysics Science Archive Research Center (HEASARC). The closest gamma-ray source to this position  is 4FGL J1946.5-5402  that lies 0.58$^{\circ}$ away and has been identified as  the millisecond pulsar PSR J1946-5403 \cite{camilo,4fgldr3}. Gamma-ray pulsars are characterized by a very sharp exponential cutoff at energies around $\sim$ 5 GeV, which suppresses any possibility of TeV emission \cite{2pc}. Furthermore, we found no cross-correlation between our $\geq$ 1 TeV sample and gamma-ray pulsars in the 4FGL-DR3 catalog. GRB 980425/SN 1998bw is located near the edge of the so-called southern Fermi bubble, but the measured gamma-ray spectrum of the bubble structure has an energy cutoff at $110 \pm 50$ GeV at high Galactic latitudes \citep{su,bubbles}. Using the spatial boundaries for the  bubbles derived by \citep{bubbles}, we estimated an $\sim$ 640 deg$^{2}$ spatial extension for each bubble. If the 49 unmatched photons were distributed isotropically  over 1.7$\pi$ steradians, we would expect roughly 2 events to fall on top of each bubble at Galactic latitudes $b \geq \vert 25^{\circ} \vert$. There are indeed 2 unmatched TeV photons in the northern bubble, but there are 5 unmatched TeV photons in the southern bubble near where GRB 980425/SN 1998bw is located. The latter is not statistically significant but it faintly hints at a possible contribution from secondary emission due to GRB 980425/SN 1998bw. Additional  LAT coverage over the next few years is needed to strengthen the case.

Within 0.15$^{\circ}$ of the 1.4-TeV photon position, there are two catalogued radio sources (8.1 and 36.3 millijansky respectively) detected as part of the Sydney University Molonglo Sky Survey (SUMSS) carried out at 843 megahertz (MHz) with the Molonglo Observatory Synthesis Telescope \cite{mauch}. In agreement, two unrelated TeV-quiet elliptical and/or lenticular galaxies with radio emission are expected within a 0.15-degree circular region based on the 843-MHz source density of 31 $\pm$ 3 sources deg$^{-2}$ \cite{mauch}.  In addition, we found no evidence for
 ROSAT, Neil Gehrels Swift Observatory, XMM-Newton, or Chandra X-ray sources 
 within 0.15$^{\circ}$ of the TeV position.

Our observational findings appear to align remarkably well with the angular size and time delay predicted for secondary gamma-ray emission produced by the propagation of UHECRs \cite{waxman1,ferrigno2005high,armengaud2006secondary,gabici2007gamma}. After travelling through the host galaxy and the IGMF, parent UHECR particles with energy $E_{p}$ can produce 
electron-positron pairs  and charged/neutral pions. Neutral pions can decay and produce ultra-energetic gamma photons with typical energy $E_{\gamma} \sim 0.1 E_{p}$. Further development of the electromagnetic cascade eventually converts most of the secondary particle energy into TeV emission. The total arrival time delay of the secondary emission $\tau$ would be the sum of the time it takes for UHECRs to cross the host galaxy $\tau_{cross}$ plus the UHECR travel time before starting the cascade $\tau_{travel}$ plus the time it takes for the cascade to generate GeV-TeV photons $\tau_{cascade}$, which can be written
\begin{equation}
 \tau =  \tau_{cross} + \tau_{travel} + \tau_{cascade}
\end{equation}

The crossing time for a $10^{20}$ eV UHECR moving through $\sim$ kpc scales and $\mu$G magnetic field within the host galaxy is given by
\begin{equation}
 \tau_{cross} \simeq 4 B_{\mu G}^{2}~{\rm days} 
\end{equation}

Once the UHECR leaves the host galaxy, it will encounter the IGMF. The maximum travel time $\tau_{travel}$ before starting a cascade in the IGMF of strength $B$ for an UHECR of energy $10^{20}$ eV located at a distance of 36.9 Mpc can be written in pG ($10^{-12}$ G) units as

\begin{equation}
\tau_{travel} \simeq B^{2}_{pG}~{\rm yrs}
\end{equation}

After the cascades start, electron pairs produced in the cascade are deflected by an IGMF of strength $B$, the magnetic deflection would induce a time delay $\tau_{cascade}$ of the secondary photons 
bracketed by the expected high-/low-energy ends of the cascade \cite{waxman1}

\begin{equation}
 \tau_{cascade} \simeq (10-1000) B^{2}_{pG}~{\rm yrs} 
\end{equation}

\noindent
For the timescales observed here $\tau_{crossing}$ is negligible and the dominant factors $\tau_{travel}$ and $\tau_{cascade}$ are nearly of the same order. 
Assuming a time delay $\tau \simeq 20.5$~yrs, this would imply that 
the IGMF strength is  of the order $B \simeq 10^{-12}$ -- 10$^{-13}$ Gauss. 
Apart from delaying secondary photons, the  IGMF may delay and spread the UHECR arrivals from GRB/SNe by millions of years \cite{waxman1}. Stellar environments where GRB/SNe take place are also a natural cradle of heavy nuclei  that would  tend to distort  UHECR trajectories \cite{piran}. GRB 980425/SN 1998bw appears to be a very fortunate case where the  time delay of the secondary TeV emission is in multi-decadal timescales and its relatively small magnetic deflection points back to the position of its source.

A GRB/SN population model that generates most detected UHECRs requires sources to lie within 200 Mpc in order to satisfy the so-called  Greisen–Zatsepin–Kuzmin (GZK) distance horizon \cite{greisen,zatsepin}. A local GRB/SN rate density  of 700 $\pm$ 360 Gpc$^{-3}$ yr$^{-1}$ has been estimated within 155 Mpc \cite{2007MNRAS.382L..21C}, which would supply a large number of UHECR accelerators within the GZK limit. The  
time delay for UHECRs propagating a distance D in a magnetic field B is given by 
$\tau(E) \simeq 10^3 {\rm yr}
(D/100{\rm Mpc})^2\, (B/10^{-11}\, {\rm G})^2\,
(E/10^{20}{\rm eV})^{-2} $ \citep{1996ApJ...462L..59M}. Super  
accelerated UHECRs with $10^{21}$ eV energies could reach us in a few decades, while their lower-energy ($10^{19}$ eV) counterparts might take over millions of years. 
If UHECR time delays and arrivals from any single GRB/SN span a million years, at any moment there could be as many as $\sim 2\times 10^{7}$ nearby GRB/SNe showering UHECRs, secondary photons and neutrinos within 200 Mpc.  In order to meet the estimated 
energy input for UHECRs  $\Gamma_{inj}$= (0.7--20) $\times$ 10$^{44}$ erg Mpc$^{-3}$ yr$^{-1}$ \cite{berezinsky2008propagation,2011NatCo...2..175C} with said GRB rate density, a total of  
 (1–30) $\times 10^{50}$ ergs per explosion are required. This seems well within reach of the energy budget inferred for GRB 980425/SN 1998bw of 2.2 $\times 10^{52}$ ergs \cite{1999ApJ...516..788W}.

An interesting outcome of such model is that it opens up the possibility that an important fraction of extragalactic background  gamma rays are due to secondary emission from magnetically deflected UHECRs. Assuming 10 matches with known gamma-ray sources, we estimated that 
49 out of 59 LAT TeV photons are unaccounted for  or 83\%. We propose that a number of these unmatched TeV photons are also secondary photons from UHECRs accelerated in nearby GRB/SNe. The only distance limit requirement here is the mean free path of the TeV photon before interacting with extragalactic background light (EBL)  photons \cite{PhysRev.155.1404}. The  best observational limits set a maximum distance for TeV emission somewhere between 0.9 and 31 Gpc \cite{2011MNRAS.410.2556D}. This would provide an even larger volume over which the Fermi LAT would be able to detect extragalactic background $\geq 1$ TeV photons.

\section{Conclusions}

A detection of  delayed TeV emission is consistent with UHECR acceleration taking place in 
GRB 980425/SN 1998bw  at a distance of 36.9 Mpc. If this interpretation is correct, GRB/SNe within the GZK sphere  might be responsible for most UHECRs, extragalactic background TeV photons, and even possibly neutrinos that are observed with current detectors. 
Further progress will result from continued observations of GRB 980425/SN 1998bw and other similar nearby events. 
A testable prediction is that  delayed secondary GeV/TeV emission and UHECRs from GRB 980425/SN 1998bw may continue to arrive in the next tens to hundreds of years.
Another testing site of the explanation presented here is the future detection of delayed GeV-TeV photons from the exceptionally bright GRB  221009A at redshift z = 0.1505 \citep{pillera}. 

In general, secondary GeV-TeV emission is expected to arrive much earlier than UHECRs \cite{waxman1}.
Based on our interpretation, extended gamma-ray, 
UHECR, and neutrino excesses may be detected around  old, nearby GRB/SN explosion sites.  Existing and future multi-messenger facilities are encouraged to monitor GRB 980425/SN 1998bw and other GRB/SNe within 200 Mpc. 
Going forward, the extraordinary wide field of view and cadence of the Fermi LAT offers unrivaled survey capabilities for additional delayed GeV-TeV emission searches.

\acknowledgments
The material is based upon work supported by NASA under award number 80GSFC21M0002.
I thank the referee for useful comments that helped improve the paper. I would like to thank Jules P. Halpern for finding the time to read the first version of the paper. 
This research has made use of data and/or software provided by the High Energy Astrophysics Science Archive Research Center (HEASARC), which is a service of the Astrophysics Science Division at NASA/GSFC. This work also used  Astropy:\footnote{http://www.astropy.org} a community-developed core Python package and an ecosystem of tools and resources for astronomy.
All data used in this paper are publicly available.
Fermi LAT data is made publicly available from the Fermi Science Support Center website immediately after processing at the following website

\url{https://fermi.gsfc.nasa.gov/ssc/data/access/} 

\noindent
HEASARC data can be accessed at the following website

\url{https://heasarc.gsfc.nasa.gov/docs/archive.html}

\bibliographystyle{JHEP.bst}
\bibliography{bibliography.bib}

\end{document}